\documentclass[lettersize,journal]{IEEEtran}
\usepackage{amsmath,amsfonts}
\usepackage{algorithmic}
\usepackage{algorithm}
\usepackage{array}
\usepackage[caption=false,font=normalsize,labelfont=sf,textfont=sf]{subfig}
\usepackage{textcomp}
\usepackage{stfloats}
\usepackage{url}
\usepackage{verbatim}
\usepackage{graphicx}
\usepackage{cite}
\usepackage{color}
\begin{document}

\title{Decision Transformers for Wireless Communications: A New Paradigm of Resource Management}

\author{Jie Zhang, Jun Li, Zhe Wang, Long Shi, Shi Jin, Wen Chen, and H. Vincent Poor
        \IEEEcompsocitemizethanks{
        \IEEEcompsocthanksitem \emph{Jie Zhang and Long Shi are with the School of Electronic and Optical Engineering, Nanjing University of Science and Technology, China. Jun Li and Shi Jin are with the National Mobile Communications Research Laboratory, Southeast University, China. Zhe Wang is with the School of Computer Science and Engineering, Nanjing University of Science and Technology, China. Wen Chen is with the Department of Electronic Engineering, Shanghai Jiao Tong University, China. H. Vincent Poor is with the Department of Electrical and Computer Engineering, Princeton University, USA. (Jun Li and Zhe Wang are the corresponding authors.)}
        }
        }

\maketitle

\begin{abstract}
As the next generation of mobile systems evolves, artificial intelligence (AI) is expected to deeply integrate with wireless communications for resource management in variable environments. In particular, deep reinforcement learning (DRL) is an important tool for addressing stochastic optimization issues of resource allocation. However, DRL has to start each new training process from the beginning once the state and action spaces change, causing low sample efficiency and poor generalization ability. Moreover, each DRL training process may take a large number of epochs to converge, which is unacceptable for time-sensitive scenarios. In this paper, we adopt an alternative AI technology, namely, Decision Transformer (DT), and propose a DT-based adaptive decision architecture for wireless resource management. This architecture innovates through constructing pre-trained models in the cloud and then fine-tuning personalized models at the edges. By leveraging the power of DT models learned over offline datasets, the proposed architecture is expected to achieve rapid convergence with many fewer training epochs and higher performance in new scenarios with different state and action spaces, compared with DRL. We then design DT frameworks for two typical communication scenarios: intelligent reflecting surfaces-aided communications and unmanned aerial vehicle-aided mobile edge computing. Simulations demonstrate that the proposed DT frameworks achieve over $3$-$6$ times speedup in convergence and better performance relative to the classic DRL method, namely, proximal policy optimization.
\end{abstract}


\section{Introduction}
\IEEEPARstart{T}{he} sixth generation of mobile communication systems (6G) is undergoing extensive development with anticipated advances in rapid data transmission, extensive coverage areas, and diverse service offerings~\cite{6G_RoadMap}. However, optimizing network performance in time-varying contexts is challenging, especially under the dynamic network topologies, fluctuating service demands, and unpredictable interference~\cite{Resource_Allocation_for_5G}. Traditional optimization methods, such as convex optimization and dynamic programming, often fall short when dealing with sequential decision making problems in the stochastic and unknown environment. Reinforcement learning (RL), as a key technology of artificial intelligence, learns the optimal policy to maximize the long-term cumulative rewards through trial-and-error without the prior modeling of the complex and dynamic environment~\cite{RL_for_Wireless_Survey}.

In recent years, extensive research has focused on wireless resource management using RL techniques. For instance, Liu et al. \cite{DBLP:journals/tvt/LiuSSLDS20} proposed a double Q-network algorithm that jointly optimizes the trajectory of unmanned aerial vehicles (UAVs) and task offloading to maximize the long-term throughput of the mobile edge computing (MEC) system. Zhang et al. \cite{IRS_JieZhang} developed a multi-agent RL framework that jointly optimizes the transmit beamforming at the base stations (BSs) and the phase shifts at the intelligent reflecting surfaces (IRSs) to maximize the long-term data rate. Additionally, Yin et al. \cite{RL_ziyan_Slicing} proposed a federated RL algorithm to optimize the time-frequency resource allocation for improving the quality of service (QoS) of the mobile users (MUs).

However, employing RL directly in wireless communications faces several challenges. Firstly, most model-free RL solutions may suffer from error propagation and instability due to the temporal difference update via the Bellman function. For example, sparse rewards or distracting signals can slow down the learning process and make it difficult to converge to near-optimal policies~\cite{DBLP:journals/tvt/Ke0DG020,RL_communication}. Second, RL algorithms often face difficulties in generalization. An RL policy well-trained in one scenario may fail to generalize to similar scenarios if the state or action space varies. For example, in a UAV-aided MEC network, changing the number of the UAVs from two to three generates a different state space. In this case, the RL-based policy learned for the 2-UAV scenario cannot be directly transferred to the 3-UAV scenario, and a new policy should be learned from scratch. This retraining process wastes additional training time and computation resources.

To address the challenges raised by RL's inefficiency in utilizing limited samples and adapting to new scenarios, the development of Decision Transformer (DT) represents a significant advancement~\cite{decision_transformer}. Drawing inspiration from the Transformer architecture~\cite{attention_is_all_your_need}, DT leverages the comprehension and generalization properties of Transformers for handling sequential decision-making problems. Unlike RL methods that rely on value function approximation and policy optimization, DT incorporates a novel approach to sequential decision-making~\cite{OfflineRL_sequential_problem, MultiGameDT}. A DT exhibits the capacity to generalize from a pre-trained model and swiftly adapt to new scenarios through local fine-tuning. DT's ability to learn from offline samples is beneficial for enhanced comprehension of intricate communication environments. Moreover, its few-shot learning capability during personalized fine-tuning allows for rapid strategy adjustments, significantly reducing training time for new scenarios~\cite{onlineDT}. This property makes DT efficiently solve policy optimization for a set of similar scenarios, particularly when there are slight variations in state or action spaces among these scenarios. Such adaptability is crucial for wireless networks with stringent real-time requirements, as it aligns with low-latency demands by facilitating online adjustments without the necessity for extensive retraining. 

In this paper, we introduce the use of DT to the field of wireless communications. The main contributions of this article are summarized as follows:
\begin{itemize}
    \item To our best knowledge, this paper is the first of its kind to develop DT-based strategies for wireless resource management. We propose a cloud-edge collaborative architecture in communication systems, where DT models are pre-trained in the cloud based on collected samples from edges and then fine-turned locally at edges to fit new scenarios. This architecture is able to efficiently facilitate model training and inferring across edges and cloud, inspiring the design of native artificial intelligence~(AI) in 6G systems.
    \item We develop two potential applications of DTs in wireless resource management, i.e., IRS-aided communications and UAV-aided MEC. For the IRS scenario, we propose an action embedding strategy to facilitate the process of pre-training and local fine-tuning of the DT model, enabling it to swiftly generalize to new scenarios where the number of IRS elements varies. For the UAV scenario, we design a parameter light-weighting method to ensure low latency and real-time response for the UAV decision model. Also, the locally fine-tuned models are adaptive to scenarios with different numbers of UAVs.
    \item Our simulations demonstrate the superiority of the proposed DT architecture over proximal policy optimization (PPO), a classic RL method. Specifically, in the IRS scenario, the convergence speed of DT is improved by $3$-$6$ times, and the performance is increased by $5.1\%$, while in the UAV scenario, the convergence speed of DT is enhanced by $4$-$6$ times, compared to PPO.
\end{itemize}

This paper is organized as follows. In Section II, we provide some background on DTs. In Section III, we propose a novel DT architecture for wireless networks with cloud-edge collaboration. In Section IV, we discuss the design of DT model for IRS and UAV scenarios. In Section V, we analyze the simulation results, while Section VI summarizes the paper.

\section{Preliminary}

Sequential decision-making in complex and dynamic environments is often challenging. Traditional optimization techniques, e.g., dynamic programming, usually assume prior and perfect knowledge of the exact mathematical modeling of the Markov decision process (MDP), in the form of state transition dynamics. RL is a model-free optimization technique for maximizing the expected return through trial and error, where return is the cumulative sum of rewards collected within an episode. An RL-based agent actively interacts with the environment to learn the optimal policies by taking actions and observing the reward and new state. Most of the model-free RL solutions adopt the temporal difference (TD) method by updating the policy via the Bellman function, which may suffer from error propagation and unstable learning, especially under sparse and distracting reward signals. Moreover, an RL policy trained in one scenario might not promise a satisfactory return in another scenario. However, retraining a new RL policy from scratch is usually time and energy consuming.

Different from the RL solutions that require extensive online interactions with the environment, DT operates on an offline dataset, modeling the joint distribution of states, actions, and rewards in an autoregressive manner. In DT's architecture, sequential data samples comprising states, actions, and returns-to-go are fed into the transformer, where the returns-to-go represents the future desired return from a given state to the end of an episode. Conditioned on the desired return and the current state, the transformer architecture with causal self-attention masks is utilized to predict the future actions via sequence generation. DT is subsequently trained by minimizing the loss between the predicted actions and the actions from the sample sequence via supervised learning. During the evaluation, the well-trained DT model can predict the near-optimal actions given the real-time states in online decision-making. 

DT has advantages over RL in terms of optimization efficiency and policy generalization capability. In cases of sparse rewards or distracting signals, the TD-based RL methods may trap into suboptimal solutions or even struggle to converge due to the slow Bellman backpropagation. In contrast, DT establishes the state-return associations through the similarity of query and key vectors within the transformer layer, which characterizes the long-range dependencies in sequential data and thus improves the optimization performance. Furthermore, with the transformer-based architecture, DT can effectively transfer the policy learned from the pre-trained datasets to the unseen scenarios through few-shot learning.

\section{System Architecture}
\begin{figure*}[htbp]
    \centering
    \includegraphics[width=1.0\textwidth]{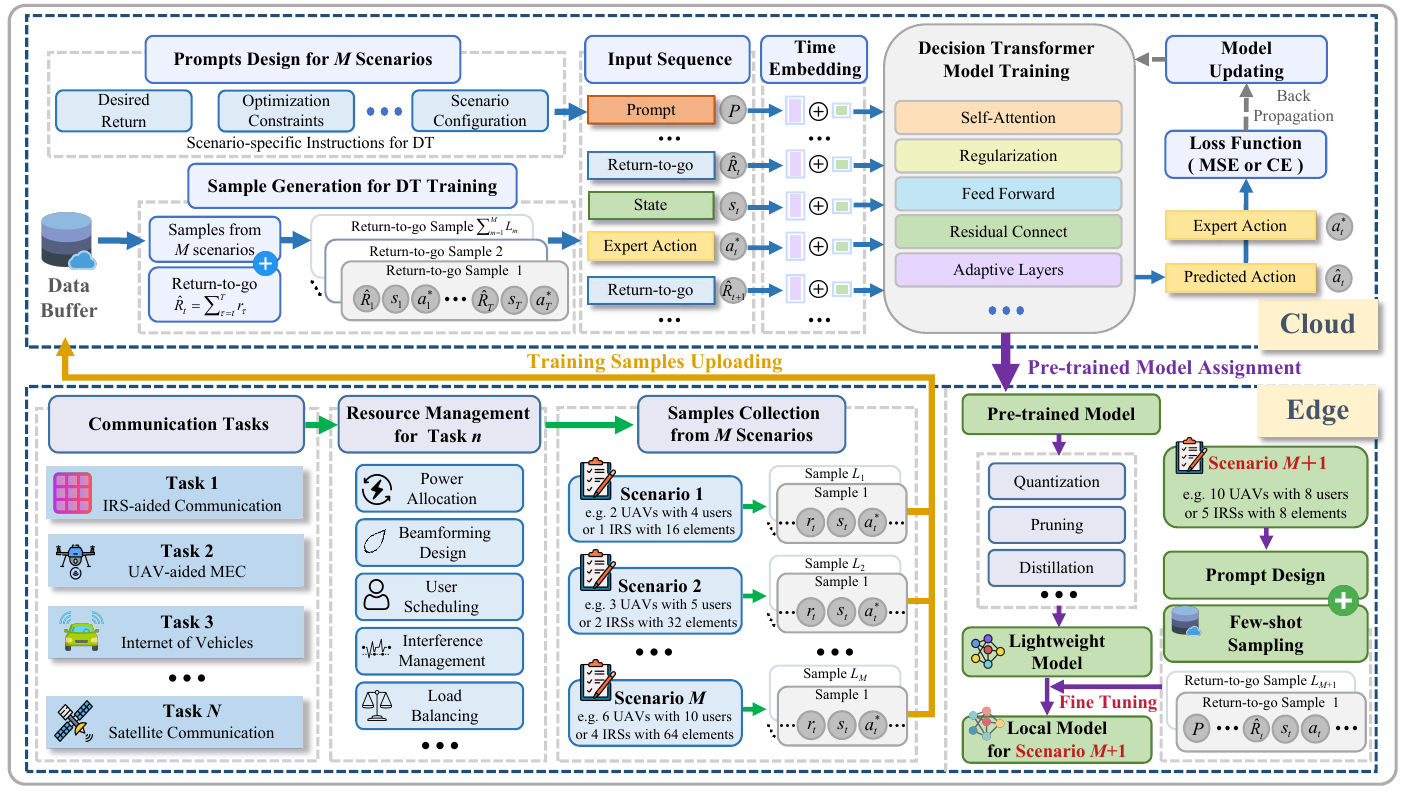}
    \caption{Cloud-edge coordinated DT architecture for wireless resource management. The platform consists of the cloud layer and the edge layer. Each communication task involves resource allocation with multiple decision variables, e.g., power allocation, beamforming design, user scheduling, interference management, and load balancing. For a specific task, samples are collected from similar scenarios and uploaded to a cloud-based data buffer. These samples undergo return-to-go processing, providing training data for the DT model. Scenario-specific prompts are designed to offer tailored instructions. The time-embedded samples are fed into the DT model, which updates its parameters by minimizing the loss between predicted and target actions. The pre-trained model is assigned to the edge in a lightweight manner. Each new scenario requires only a few samples for fine-tuning to establish a personalized local model.}
    \label{DT architecture}
\end{figure*}

In this section, we construct a cloud-edge collaborative architecture of wireless resource management with a DT. As shown in Fig.~\ref{DT architecture}, it encompasses both cloud and edge workflows, supporting a variety of wireless communication tasks, e.g., IRS-aided communications~\cite{IRS_survey}, UAV-aided MEC~\cite{DBLP:journals/tvt/LiuSSLDS20}, the Internet of Vehicles (IoV), and satellite communications. For each communication task, it is crucial to optimize the resource allocation for enhancing the long-term system performance, e.g., maximizing the network throughput or reliability, or minimizing the delay or costs, under the system constraints. For example, resource management may include, but is not limited to, power control, computation offloading, multi-antenna beamforming, user scheduling, or interference management.

\subsection{Sample Collection and Uploading}

For a specific communication task, there may exist multiple scenarios (i.e., Scenarios $1$ to $M$) with similar optimization problems but different settings. Taking the UAV-aided MEC network as an example, the objective in Scenario $1$ is to jointly optimize the trajectories of $2$ UAVs to serve $4$ users, aiming to maximize the long-term throughput, whereas Scenario $2$ might involve $3$ UAVs serving $5$ users with the same objective. Taking an IRS-aided communication task as another example, the aim for Scenario $1$ is to jointly optimize the beamforming design of a multi-antenna BS and a passive IRS with $16$ elements in order to maximize the average rate at an MU, while Scenario $2$ might involve $2$ IRSs with $32$ elements to achieve the same goal. To solve these optimization problems, we can adopt sequential optimization methods, e.g., dynamic programming or RL, to obtain near-optimal policies. However, these policies often fail to adapt to unseen environments, and the retraining is time-consuming and computation-expensive. In our structure, high-quality samples obtained from these policies are collected at discrete time steps across the sequences of length $T$, encompassing state $s_{t}$, expert action $a_{t}^{*}$, and reward $r_{t}$ for each time step $t$. Specifically, samples of expert actions can be generated via the near-optimal policies obtained through the well-trained RL or heuristic algorithms. These samples from different edge scenarios are then uploaded to a data buffer at the cloud for the DT model training.

\subsection{Model Pre-training and Assignment}
The cloud serves as a centralized platform for developing general DT models tailored to each communication task. Rewards from samples in the data buffer are restructured into the return-to-go, $\hat{R}_{t}=\sum_{\tau=t}^{T}r_{\tau}$, representing the future desired returns for each time step $t$. Moreover, we adopt a prompt design module to provide the specific-scenario instructions for the model, which include desired returns, resource constraints, and environment configurations. Next, the samples are time-embedded and then sequentially fed with the prompt into the DT model. The DT model employs a transformer-based structure that leverages the self-attention mechanisms for effective sequence generation and captures the long-range temporal relationships within the data. The integration of regularization layers, feedforward networks, and residual connections enhances the learning process and mitigates overfitting. Additionally, adaptive layers enable the DT model to adapt its parameters to the new scenarios. The DT model is trained by minimizing the loss function between its output (predicted action $\hat{a}_{t}$) and the expert action $a_{t}^{*}$ from the training sequence, where the loss can take the form of mean squared error (MSE) or cross-entropy (CE). Through this supervised learning approach, the pre-trained DT model can generate a general policy for a specific task.


\subsection{Model Fine-tuning and Generalization}
The general DT model pre-trained on the cloud is usually in large size, which may not be directly feasible for edge devices with limited computational resources and storage capacity. It is thus crucial to lightweight the pre-trained model at the edge via compression methods such as quantization, pruning, and distillation. These techniques reduce the model’s complexity and size, making it suitable for deployment on edge devices. For a new scenario $M+1$, the number of IRS elements or the number of UAVs may be different from the pre-trained $M$ scenarios, where retraining the policies from scratch is costly. Instead, we can fine-tune the lightweight model with few-shot samples. For sample collection, the lightweight DT can be used for directly interacting with the environment in the new scenario to generate samples. Besides, the samples of non-expert actions generated during the training process of the RL or heuristic algorithms can also be collected. There are various methods for the fine-tuning process, such as layer freezing, prefix-tuning, and low-rank adaptation. Take the layer freezing technique as an example, the weights of the lower layers are fixed to preserve the general features learned from the offline datasets, while the output layer is adjusted using the few-shot newly collected samples to ensure the personalized DT model adapts effectively to the new scenario.
    
Through the above cloud-based pre-training and edge-based fine-tuning processes, the DT-based cloud-edge collaborative framework ensures rapid policy generalization across different scenarios in the dynamic communication systems.

\section{Two Typical Application Tasks}
In this section, we specify the applications of the DT architectures in two typical tasks, i.e., IRS-aided communications and UAV-aided MEC.

\subsection{Task 1: IRS-aided Communications}
An IRS is a passive device that consists of multiple adjustable elements, each capable of independently controlling the phase and amplitude of reflecting signals~\cite{IRS_survey}. We first take the IRS-aided communication system as an example, where the IRS enhances channel quality between BSs and users by adjusting the phase shits of its elements.

\begin{figure}[htbp]
    \centering
    \includegraphics[width=0.5\textwidth]{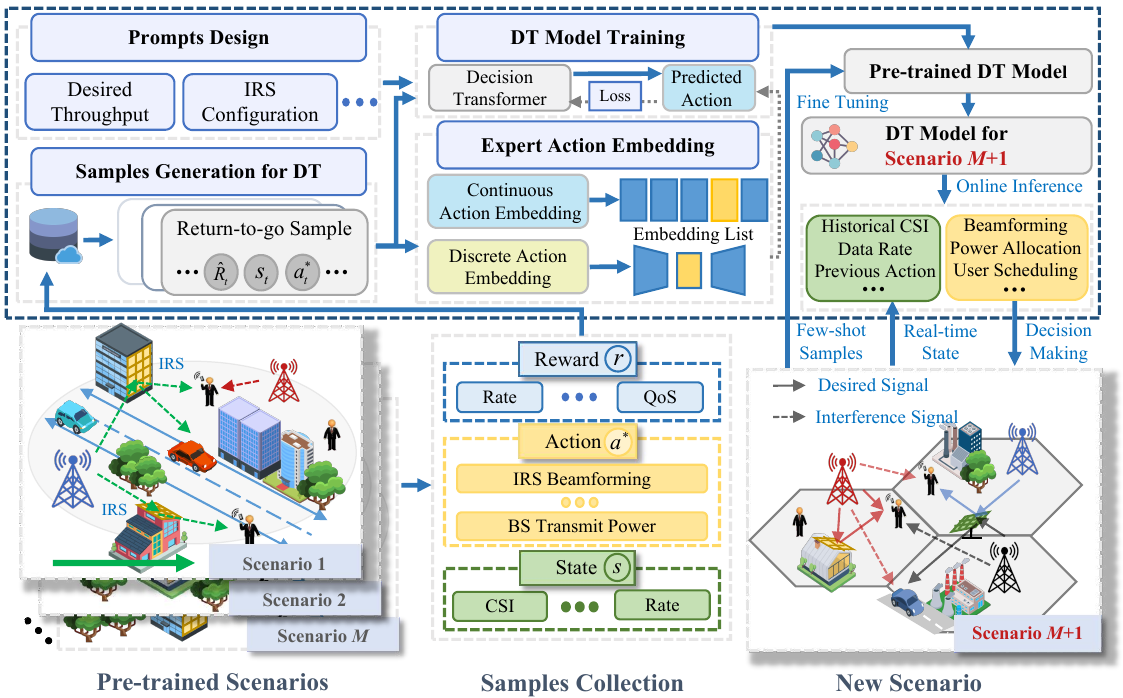}
    \caption{DT design for IRS-aided communications. The DT model is pre-trained with multiple scenarios to learn the policies of BS's power control and IRS's beamforming. The model trains with scenario-specific prompts and an action embedding technique for managing both discrete and continuous actions. In real-time deployment, the model swiftly adapts to the new scenario by fine-tuning the pre-trained DT model with few-shot learning.}
    \label{IRS DT architecture}
\end{figure}

As illustrated in Fig.~\ref{IRS DT architecture}, we propose a DT architecture for IRS-assisted communications, aiming to maximize the expected cumulative throughput by jointly optimizing the IRS's beamforming vectors and the BS's transmit power. Firstly, the DT model collects the training samples from multiple historical scenarios, which include states, actions, and rewards. For example, we can take the channel state information (CSI) and the feedback of the data rate in the previous step as the states, the transmit power of the BS and the beamforming vector of the IRS as the actions, and the immediate data rate and QoS as the rewards. Consider multiple scenarios that include various objectives (e.g., desired throughputs) and different constraints (e.g., heterogeneity in BS maximum power and IRS configurations). The above information serves as the prompt for the DT model and provides scenario-specific guidance during training. By integrating this richer context, the DT model can make more rational decisions that fully consider the unique characteristics of each scenario.

Regarding the action space, we confront the challenge of a hybrid action space comprising both discrete and continuous actions, such as the IRS's phase shifts and the BS's transmit power. The direct application of DT's outputs to such a mixed action space will potentially lead to suboptimal policies due to the mismatch in action representations. To resolve this issue, we propose an action embedding technique that maps actions into an embedding space. For the discrete actions, we employ the nearest neighbor matching method in the embedding space to find the closest viable discrete action. For the continuous actions, an auto-regressive method is used for embedding, enhancing the DT model's ability to generate continuous actions. This innovation not only addresses the inherent challenges of a mixed action space but also enhances the model's performance.

After undergoing the model training, the pre-trained DT model needs to be deployed to fit the new edge scenarios by fine-tuning with few-shot samples. During the evaluation, the DT model can utilize these real-time inputs in the new scenario, e.g., channel state, data rate feedback, desired throughput, and constraint prompts, to swiftly generate high-quality actions of power control and beamforming.

\subsection{Task 2: UAV-aided MEC}
We take UAV-assisted MEC as another example, where the UAVs act as aerial servers to jointly deliver the computation offloading services to multiple ground MUs. Our objective is to maximize the expected cumulative rate in the system by jointly optimizing the UAVs’ flight paths, workload scheduling, and user association. Due to the heterogeneity in the UAVs' limited resources, e.g., computational capabilities, service ranges, and battery capacities in different scenarios, it is necessary to lightweight the pre-trained DT model and fine-tune it at the UAV edge locally.

\begin{figure}[htbp]
    \centering
    \includegraphics[width=0.48\textwidth]{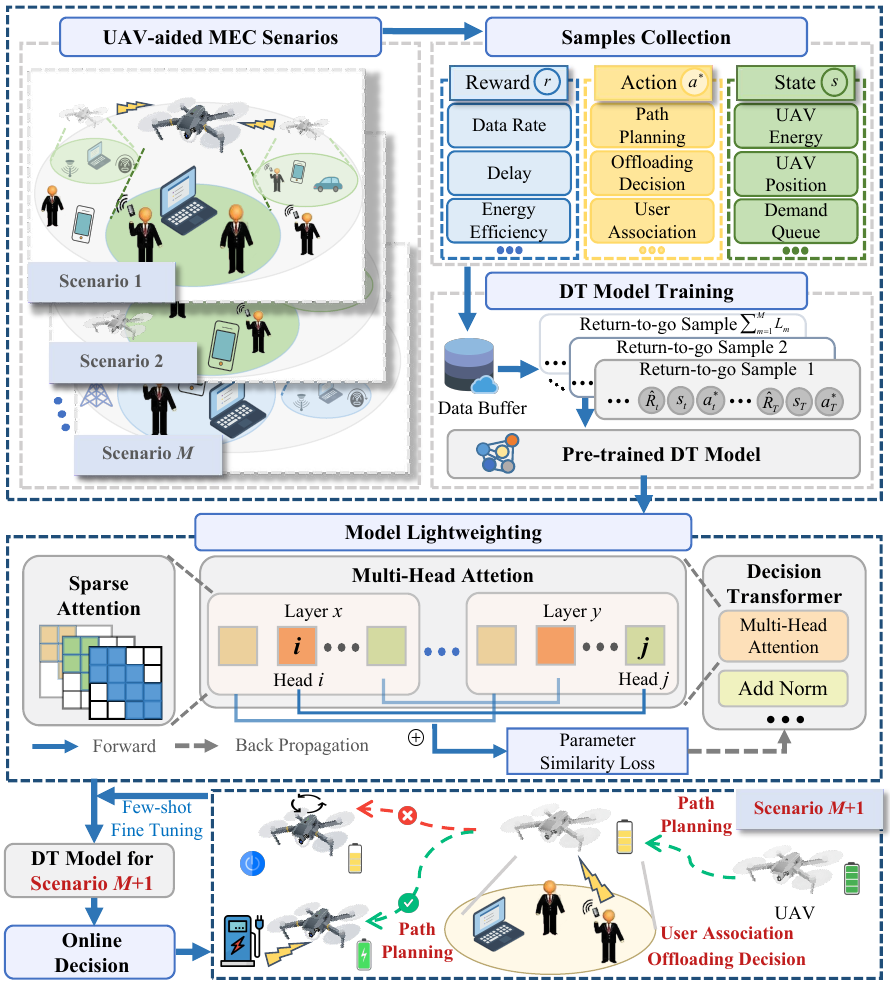}
    \caption{The DT-based resource management for UAV-aided MEC with model lightweighting. This architecture incorporates parameter sharing and sparse attention mechanisms to reduce the neural network's complexity and computational demands. Samples are collected from multiple UAV scenarios to pre-train the DT model. For online deployment, the lightweight DT model is fine-tuned to fit the new scenario.
    }
    \label{UAV DT architecture}
\end{figure}

As shown in Fig.~\ref{UAV DT architecture}, we first establish the dataset by collecting training samples from the existing UAV-aided MEC scenarios. In these scenarios, we take the UAVs' remaining energy, locations, and demand queues as the states, the path planning, offloading decision, and user association as the actions, and the data rates, latency, or energy efficiency as the rewards. The design of prompts for UAV scenarios incorporates constraints that include the battery capacity of each UAV, the computational capabilities of the UAVs, and the number of users and UAVs within the scenario. To address the critical latency and responsiveness requirements in UAV scenarios, we have developed a lightweighting strategy for our DT architecture. We employ both parameter sharing and sparse attention mechanisms within the transformer structure. Initially, we introduce parameter sharing across different heads, allowing them to utilize a common set of weights rather than learning a unique transformation at each head. This approach significantly reduces the total number of neural network parameters and decreases the model complexity. Additionally, we replace the conventional dense attention mechanism by implementing a sparse attention method. This mechanism selectively masks future information and limits attention weight calculations to a pre-defined window, focusing primarily on the past information deemed more relevant for UAV tasks. Such a strategy reduces the computational complexity of attention calculations from quadratic to linear with respect to the sequence length, thus considerably reducing computational overhead. Furthermore, to balance the trade-off between the inference speed and performance, we also introduce a loss function based on parameter similarity. This lightweight method reduces the model's size without compromising the optimality of the policy.

When deployed in a new UAV-aided MEC scenario, the model is fine-tuned based on the real-time data. This personalized fine-tuning process enables the DT model to guide the UAVs to optimize their path planning, offloading decisions, and user association. The DT model's capability to generalize across various scenarios reduces not only the response times but also the energy costs of the resource-constrained UAVs.

\section{Numerical Results and Analysis}

In this section, we present detailed simulations of two applications to evaluate the performance of our proposed DT architecture. The DT model is pre-trained and fine-tuned on NVIDIA RTX A6000 GPU hardware, employing the AdamW optimizer with a learning rate of $10^{-4}$. The main architecture encompasses three transformer blocks, each designed with a causal attention network, a feedforward network, and a dropout network with a dropout rate of $0.1$. The training process is conducted with a batch size of $64$.
\begin{figure}[htbp]
    \centering
    \includegraphics[width=0.5\textwidth]{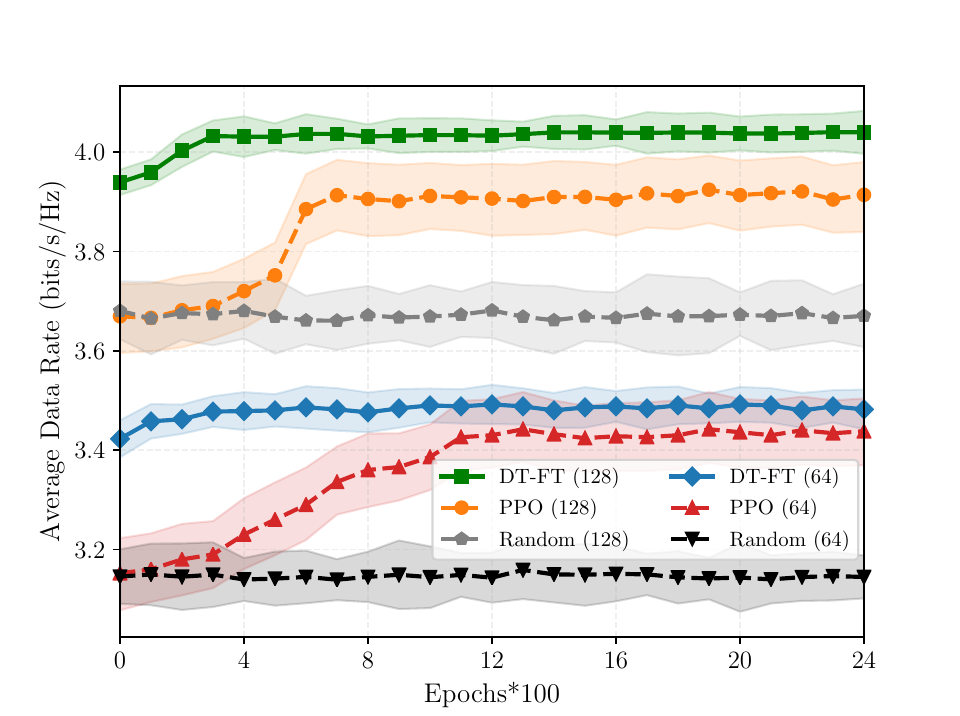}
    \caption{Performance and convergence speed comparison between DT, PPO and random methods in IRS-aided communications. We consider two scenarios with $128$ and $64$ IRS elements, respectively.} 
    \label{IRS simulation}
\end{figure}

We first consider the scenarios of an IRS-aided communication system consisting of a BS, an IRS, and a MU. The channel model in~\cite{IRS_JieZhang} is adopted. Each training epoch consists of $100$ time slots. Each scenario exhibits channel characteristics that possess slightly distinct channel exponents. We define the CSI as the state, the IRS beamforming as the action, and the data rate as the reward. For pre-training, we utilize offline datasets collected via the PPO algorithm~\cite{PPO_algorithm}. During the fine-tuning process, a limited number of samples are generated through direct interaction with the environment in the new scenario using the PPO algorithm from scratch. These few-shot samples are utilized to fine-tune the DT model, namely DT-FT. Then we evaluate the performance of the DT-FT in the new scenario.

In Fig.~\ref{IRS simulation}, we compare the average data rate performance of the proposed DT-FT method with the PPO algorithm and a random selection method by considering $128$ and $64$ IRS elements as the two new scenarios. The comparison distinctly illustrates that the performance of the DT-FT for the two new scenarios is superior to that of the PPO. To be specific, the DT model achieves $5.1$ percent performance improvement and exhibits a convergence speed that is $3$-$6$ times faster than that of the PPO algorithm. This superior performance is primarily attributed to DT's sequence learning capability, which enables it to effectively capture the intricate inner relationships between states, actions, and rewards. Moreover, the DT's proficiency in leveraging offline training samples allows it to swiftly converge in new scenarios with only a limited number of samples required for fine-tuning.

We further consider the scenarios of UAV-aided MEC, where UAVs fly within a $100\times 100$ $\mathrm{m}^{2}$ region to provide computing services to ground users. Our objective is to optimize the flight trajectories of the UAVs and their associations with users to maximize the long-term throughput of the computing workload. The simulation encompasses $10$ users, whose mobility is characterized by a Gaussian Markov model. The volumes of computational workload are uniformly distributed within the range of $[10,20]$ Mb. The state is composed of the UAVs' positional coordinates and the remaining volume of workload. The action includes the UAVs' path planning and their selections of users to serve. The reward is defined as workload throughput. The PPO algorithm is utilized for collecting training samples that are further used to pre-train the DT model. Subsequently, the DT model undergoes a fine-tuning process specifically tailored to the new scenarios.

\begin{figure}[htbp]
    \centering
    \includegraphics[width=0.5\textwidth]{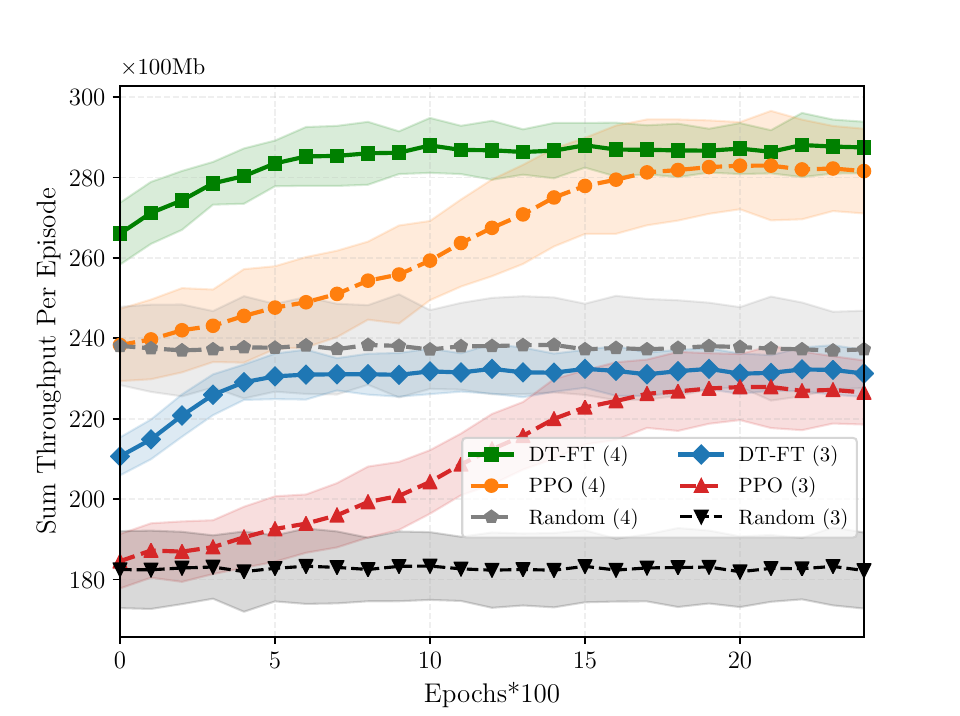}
    \caption{Performance and convergence speed comparison between DT, PPO and random methods in UAV-aided MEC. We consider two scenarios with $3$ UAVs and $4$ UAVs, respectively.} 
    \label{UAV simulation}
\end{figure}

As depicted in Fig.~\ref{UAV simulation}, we plot the sum throughput per episode utilizing two new scenarios with $3$ UAVs and $4$ UAVs, respectively. The results from the new scenarios show the superiority of DT in terms of rapid convergence compared to the PPO, which is $4$-$6$ times faster, showcasing DT's powerful few-shot learning capability. This significant advantage highlights the proficiency of DT in distilling generalized knowledge from pre-trained scenarios and effectively adapting to new scenarios, indicating its significant potential and applicability in the UAV-aided MEC system.

\section{Conclusions}
In this paper, we have explored the DT-based stochastic resource management for wireless communications, addressing the generalization challenge that is commonly encountered in conventional RL. We have proposed a novel DT architecture for leveraging its strengths based on a cloud-edge coordination manner. Simulations indicate that the proposed DT architecture is superior to conventional RL in terms of convergence speed and performance, with $4$-$6$ times faster and $5.1$ percent performance improvement. Our approach is not limited to only the discussed of IRS-aided communications and UAV-aided MEC, but is also a promising approach to developing scalable solutions for other scenarios, such as resource allocations for the Internet of Things, satellite communications, and so on.

\section{Acknowledgments}
This work was supported in part by a Chinese National key project under Grant 2020YFB1807700, in part by the National Natural Science Foundation of China (NSFC) under Grants 62071296, 62202232, 62371239, and 62471204, in part by the U.S. National Science Foundation under Grant ECCS-2335876, in part by the Key Technologies R$\&$D Program of Jiangsu (Prospective and Key Technologies for Industry) under Grants BE2023022 and BE2023022-2, in part by the Natural Science Foundation of Jiangsu Province under Grant BK20210331, in part by the Jiangsu Specially-Appointed Professor Program 2021, and in part by Shanghai Kewei under Grant 22JC1404000.

\bibliographystyle{IEEEtran}
\bibliography{IEEEabrv,Ref}

\begin{IEEEbiographynophoto}
{Jie Zhang} (zhangjie666@njust.edu.cn) received the B.S. degree from the School of Electronic and Optical Engineering, Nanjing University of Science and Technology, Nanjing, China, in 2019, where he is working toward the Ph.D. degree currently. His research interests include reinforcement learning, multi-agent system and intelligent reflecting surface.
\end{IEEEbiographynophoto}

\begin{IEEEbiographynophoto}
{Jun Li} [M'09, SM'16] (jleesr80@gmail.com) received the Ph.D. degree in Electronic Engineering from Shanghai Jiao Tong University, Shanghai, P.R. China in 2009. From January 2009 to June 2009, he worked in the Department of Research and Innovation, Alcatel Lucent Shanghai Bell as a Research Scientist. From June 2009 to April 2012, he was a Postdoctoral Fellow at the School of Electrical Engineering and Telecommunications, the University of New South Wales, Australia. From April 2012 to June 2015, he was a Research Fellow at the School of Electrical Engineering, the University of Sydney, Australia. From June 2015 to now, he is a Professor at the School of Electronic and Optical Engineering, Nanjing University of Science and Technology, Nanjing, China. He was a visiting professor at Princeton University from 2018 to 2019. His research interests include network information theory, game theory, distributed intelligence, multiple agent reinforcement learning, and their applications in ultra-dense wireless networks, mobile edge computing, network privacy and security, and industrial Internet of Things. He has co-authored more than 200 papers in IEEE journals and conferences, and holds one US patent and more than 10 Chinese patents in these areas. He is serving as an editor of IEEE Transactions on Wireless Communication and a TPC member for several flagship IEEE conferences.
\end{IEEEbiographynophoto}

\begin{IEEEbiographynophoto}
{Zhe Wang} [M] (zwang@njust.edu.cn) received the Ph.D. degree in electrical engineering from The University of New South Wales, Sydney, Australia, in 2014. From 2014 to 2020, she was a Research Fellow with The University of Melbourne, Australia, and Singapore University of Technology and Design, Singapore, respectively. She is currently a Professor with the School of Computer Science and Engineering, Nanjing University of Science and Technology, Nanjing, China. Her research interests include applications of optimization, reinforcement learning, and game theory in communications and networking. She serves as an editor of IEEE Open Journal of the Communications Society.
\end{IEEEbiographynophoto}

\begin{IEEEbiographynophoto}
{Long Shi} [SM] (slong1007@gmail.com) received the Ph.D. degree in Electrical Engineering from the University of New South Wales, Sydney, Australia, in 2012. From 2013 to 2016, he was a Postdoctoral Fellow at the Institute of Network Coding, Chinese University of Hong Kong, China. From 2014 to 2017, he was a Lecturer at Nanjing University of Aeronautics and Astronautics, Nanjing, China. From 2017 to 2020, he was a Research Fellow at the Singapore University of Technology and Design. Now he is a Professor at the School of Electronic and Optical Engineering, Nanjing University of Science and Technology, Nanjing, China. His research interests include wireless communications, decentralized security, and edge intelligence. He is serving as an editor of IEEE Transactions on Cognitive Communications and Networking.
\end{IEEEbiographynophoto}

\begin{IEEEbiographynophoto}
{Shi Jin} [SM] (jinshi@seu.edu.cn) received the B.S. degree in communications engineering from the Guilin University of Electronic Technology, Guilin, China, in 1996, the M.S. degree from the Nanjing University of Posts and Telecommunications, Nanjing, China, in 2003, and the Ph.D. degree in information and communications engineering from Southeast University, Nanjing, in 2007. From June 2007 to October 2009, he was a Research Fellow with the Adastral Park Research Campus, University College London, London, U.K. He is currently affiliated with the Faculty of the National Mobile Communications Research Laboratory, Southeast University. His research interests include wireless communications, random matrix theory, and information theory. He and his coauthors were awarded the 2011 IEEE Communications Society Stephen O. Rice Prize Paper Award in the field of communication theory, the 2022 Best Paper Award, and the 2010 Young Author Best Paper Award by the IEEE Signal Processing Society. He was an Associate Editor of the IEEE Transactions on Wireless Communications, IEEE Communications Letters, and IET Communications. He serves as an Area Editor for the IEEE Transactions on Communications and IET Electronics Letters.
\end{IEEEbiographynophoto}

\begin{IEEEbiographynophoto}
{Wen Chen} [M’03, SM’11] (wenchen@sjtu.edu.cn) received the B.S. and M.S. from Wuhan University, China in 1990 and 1993 respectively, and the Ph.D. from University of Electro-communications, Japan in 1999. He is now a tenured Professor with the Department of Electronic Engineering, Shanghai Jiao Tong University, China, where he is the director of Broadband Access Network Laboratory. His research interests include multiple access, wireless AI and reconfigurable intelligent surface enabled communications. He has published more than 170 papers in IEEE journals with citations more than 10,000 in Google Scholar.
\end{IEEEbiographynophoto}

\begin{IEEEbiographynophoto}
{H. Vincent Poor} [F'87] (poor@princeton.edu) is the Michael Henry Strater University Professor at Princeton University, where his interests include information theory, machine learning and network science, and their applications in wireless networks, energy systems, and related fields. He is a member of the U.S. National Academy of Engineering and the U.S. National Academy of Sciences, and a foreign member of the Royal Society and other national and international academies. He received the IEEE Alexander Graham Bell Medal in 2017.
\end{IEEEbiographynophoto}
\end{document}